\documentclass[preprint, sort&compress, english]{elsarticle}

\usepackage{lmodern,amsmath,amssymb,amsthm,mathtools}

\usepackage[utf8]{inputenc}
\usepackage[T1]{fontenc}

\usepackage[h]{esvect}
\usepackage{graphicx}
\usepackage{placeins}
\usepackage[dvipsnames]{xcolor}
\usepackage{booktabs}
\usepackage{placeins}
\usepackage{pifont}
\usepackage{caption}

\newcommand{\cmark}{\ding{52}}%
\newcommand{\xmark}{\ding{56}}%

\usepackage{etoolbox}

\newcommand{\dd}[2][]{\ifstrempty{#1}{\mathop{\mathrm{d}#2}}{\mathop{\mathrm{d}^{#1}#2}}}

\usepackage{hyperref}
\usepackage{hypernat}

\newcommand{\Eqref}[1]{Eq.~\eqref{#1}}
\newcommand{\Eqsref}[1]{Eqs.~\eqref{#1}}
\newcommand{\Secref}[1]{Sec.~\ref{#1}}
\newcommand{\Tabref}[1]{Tab.~\ref{#1}}
\newcommand{\Figref}[1]{Fig.~\ref{#1}}
\newcommand{\Figsref}[1]{Figs.~\ref{#1}}
\newcommand{\Refcite}[1]{Ref.~\cite{#1}}

\title{A Systematic Approach to Numerical Dispersion in Maxwell
Solvers}

\author[HIJ]{Alexander Blinne}
\author[HIJ]{David Schinkel}
\author[HIJ,JenaUni]{Stephan Kuschel}
\author[HIJ]{Nina Elkina}
\author[HIJ]{Sergey G. Rykovanov}
\author[HIJ,JenaUni,BelfastUni]{Matt Zepf}

\address[HIJ]{Helmholtz Institute Jena, Fr\"{o}belstieg 3, 07743 Jena, Germany}
\address[JenaUni]{Institute for Optics- and Quantumelectronics, Friedrich Schiller University Jena, Max-Wien-Platz 1, 07743 Jena, Germany}
\address[BelfastUni]{Department of Physics and Astronomy, Queen's University Belfast, Belfast, BT7 1NN, UK}
\tnotetext[t1]{\copyright  2017. This manuscript version is made available under the CC-BY-NC-ND 4.0 license \url{http://creativecommons.org/licenses/by-nc-nd/4.0/}}

\begin{document}

\begin{abstract}
The finite-difference time-domain (FDTD) method is a well established method
for solving the time evolution of Maxwell's equations.
Unfortunately the scheme introduces numerical dispersion and therefore phase and group velocities which deviate from the correct values. The solution to Maxwell's equations in more than one dimension results in non-physical  predictions such as numerical dispersion or numerical Cherenkov radiation emitted by a relativistic electron beam propagating in vacuum.

Improved solvers, which keep the staggered Yee-type grid for electric and magnetic fields, generally modify the spatial derivative operator in the Maxwell-Faraday equation by increasing the computational stencil. These modified solvers can be characterized by different sets of coefficients, leading to different dispersion
properties. In this work we introduce a norm function to rewrite the
choice of coefficients into a minimization problem. We solve this
problem numerically and show that the minimization procedure leads to
phase and group velocities that are considerably closer to $c$ as compared to schemes with
manually set coefficients available in the literature. Depending on a specific
problem at hand (e.g. electron beam propagation in plasma,
high-order harmonic generation from plasma surfaces, etc), the norm
function can be chosen accordingly, for example, to minimize the
numerical dispersion in a certain given propagation
direction. Particle-in-cell simulations of an electron beam propagating in vacuum using our solver are provided.

\end{abstract}

\maketitle

\section{Introduction}

The finite-difference time-domain (FDTD) method is widely used for the
simulation of electromagnetic wave propagation in different scenarios ranging from the simulation of antennas to astrophysical problems~\cite{inan2011numerical}.
It is also the standard method to solve Maxwell's equations in particle-in-cell (PIC) simulations, where plasma, represented by macroparticles, and electromagnetic fields are treated in a self-consistent manner~\cite{Birdsall:1991}.
In the traditional FDTD method, also known as the Yee scheme~\cite{Yee1966}, the electromagnetic fields are located on a staggered grid providing the centered differencing using the nearest grid points. For many years Yee's scheme has been extensively used in electromagnetic plasma simulations leading to many important results.
One of the main drawbacks of Yee's scheme is the numerical dispersion which leads to a lower than $c$ phase and group velocity \cite{Valcarce2010}, particularly at higher frequencies (here, $c$ is the speed of light in vacuum). The artificial slow-down of the phase velocity means that a relativistic electron propagating along a straight line with constant velocity in free-space will emit artificial numerical Cherenkov radiation~\cite{Godfrey1974}, which is of particular interest for PIC simulations, notably in the context of Laser Plasma Accelerators (LPAs).
The deviation of the group velocity from its vacuum value may also adversely effect simulations of LPAs by modifying wave-breaking.

A lot of effort has been taken to reduce the numerical dispersion of Maxwell
solvers, especially for the LPA simulations using PIC codes~\cite{Godfrey:2014vxa}. Several approaches
have been proposed including spectral methods~\cite{Liu1997, Lifschitz2009,
Vay2013,Yu2015, Andriyash2016, Lehe2016}, filtering \cite{Greenwood2002,
Wang2003, Greenwood2004, Vay2011, Webb2016}, directional
splitting~\cite{Yanenko2012, LichtersLPIC} and modified computation stencils~\cite{Lele1992, Georgakopoulos2002, Greenwood2004, Panaretos2006, Finkelstein2007, Vay2011, Karkkainen2006, Pukhov1999, Cowan2013, Lehe2013, Nuter2014}.
In the latter methods the stencil for the computation of the spatial derivatives of the fields is extended.
For the usage in PIC simulations, it is non-trivial to find a particle current weighting scheme that conserves charge w.\,r.\,t. the extended stencils \cite{Greenwood2004}. Therefore in most cases only the stencil used for the spatial derivatives for the electric field in the Maxwell-Faraday equation is modified.
Depending on the choice of the coefficients the field solver will have different dispersion properties. For example, as shown in \Refcite{Lehe2013}, one can design a solver in which higher frequencies can propagate with phase velocities slightly higher than $c$. Consequently this leads to a significant reduction of artificial Cherenkov radiation in a chosen direction, which increases the accuracy of LPA simulations. Another set of coefficients can lead to dispersion-free propagation along the chosen axis, as shown in \Refcite{Pukhov1999}.
For the off-axis propagation the setups presented in Refs.~\cite{Lehe2013, Pukhov1999} remain dispersive.

In this paper a systematic approach to modify the computational stencil in Maxwell solvers is presented. By introducing a norm in coefficient space a minimization algorithm can automatically deduce stencil coefficients either with minimized global dispersion or with minimized dispersion in a given range of frequencies and angles. We show that our approach can lead to dispersion properties better than those of Maxwell solvers with published coefficient sets.

This paper is organized as follows. In \Secref{sec:stencils} the extended stencil method is introduced and conditions for the coefficients are discussed. In \Secref{sec:dispersion_relation} expressions for the dispersion relation are provided. In \Secref{sec:norm} we introduce the norm function, and discuss that simple numerical minimization methods can provide an optimal set of coefficients for the best (with respect to numerical dispersion) performance. Examples of optimized stencils and their properties are discussed in \Secref{sec:optimized_stencils}. In \Secref{sec:numerical} numerical PIC simulation examples are provided. Finally, \Secref{sec:conclusions} contains conclusions.

\section{Stencils for Maxwell solvers}\label{sec:stencils}
In order to solve Maxwell's equations
\begin{align}
 \vv{\nabla}\times \vv{E} &= -\partial_t \vv{B}\,,                           \label{eq:Maxwell_Faraday}\\
 \vv{\nabla}\times \vv{B} &= \frac{1}{c^2}\partial_t \vv{E} + \mu_0\vv{J}\,, \label{eq:Maxwell_Ampere} \\
 \vv{\nabla} \cdot \vv{E} &= \frac{\rho}{\varepsilon_0}     \,,              \label{eq:Maxwell_Gauss_E}\\
 \vv{\nabla} \cdot \vv{B} &= 0                                           \label{eq:Maxwell_Gauss_B}
\end{align}
on the grid, stencils for the derivative operators have to be imposed.
Our starting point is Yee's scheme from \Refcite{Yee1966} which is based on the combination of a staggered grid with second order accurate central difference stencils.
The Maxwell-Faraday equation \Eqref{eq:Maxwell_Faraday} and the Maxwell-Ampere equation \Eqref{eq:Maxwell_Ampere} are used, in a numerical form, to update the fields from one time step to the next.
The remaining equations, \Eqsref{eq:Maxwell_Gauss_E} and \eqref{eq:Maxwell_Gauss_B}, are seen as constraint equations.
They must be, numerically, conserved from step to step and they should be valid at the initial time.
In the context of a Particle-in-Cell (PIC) simulation, it is the task of the current weighting scheme to ensure that $\vv{\nabla} \cdot \vv{E}$ is not only conserved, but reflects the actual charge density given by the same moving charged particles that are the sources of $\vv{J}$.

The extended stencil scheme is introduced by replacing, in some but not all the equations, the simple central difference stencil by an extended stencil that takes additional grid points into account.
For an arbitrary field component at time $t=n\Delta t$ and position
$(x,y,z)=(i\Delta x, j\Delta y, k \Delta z)$ with $\Delta t, \Delta x, \Delta y, \Delta z$ the discretization steps in time and space, respectively, we impose the standard second order
stencil for the time derivative
\begin{gather}
\partial_t X \rightarrow D_t( X^n_{i\,j\,k})=\frac{1}{\Delta t}\left(X^{n+\frac{1}{2}}_{i\,j\,k} -
X^{n-\frac{1}{2}}_{i\,j\,k}\right).
\end{gather}
To avoid complications in the current deposition inside the PIC Codes the stencil for the Maxwell-Ampere equation \Eqref{eq:Maxwell_Ampere} is chosen as the Yee stencil and therefore remains unmodified:
\begin{gather}
\nabla_x\rightarrow D_x( X^n_{i\,j\,k})=\frac{1}{\Delta
x}\left(X^{n}_{i+\frac{1}{2}\,j\,k} - X^{n}_{i-\frac{1}{2}\,j\,k}\right)\mathrm{,}
\end{gather}
and similarly for the remaining spatial coordinates.
The stencil for the Maxwell-Faraday equation \Eqref{eq:Maxwell_Faraday} is generalized to:
\begin{equation}
\begin{aligned}
\nabla_x\rightarrow D^*_x(X^n_{i\,j\,k}) &= \frac{\alpha_x}{\Delta
x}\left(X^{n}_{i+\frac{1}{2}\, j\,k} - X^{n}_{i-\frac{1}{2}\,j\,k}\right) \\
&+\frac{\delta_x}{\Delta x}\left(X^{n}_{i+\frac{3}{2}\, j\,k} - X^{n}_{i-\frac{3}{2}\,j\,k}\right)\\
&+ \frac{\beta_{xy}}{\Delta x}\left(X^{n}_{i+\frac{1}{2}\, j+1\,k} -X^{n}_{i-\frac{1}{2}\,j+1\,k}\right) \\
&+ \frac{\beta_{xy}}{\Delta x}\left(X^{n}_{i+\frac{1}{2}\, j-1\,k} -
X^{n}_{i-\frac{1}{2}\,j-1\,k}\right) \\
&+\frac{\beta_{xz}}{\Delta x}\left(X^{n}_{i+\frac{1}{2}\, j\,k+1} -
X^{n}_{i-\frac{1}{2}\,j\,k+1}\right) \\&
+ \frac{\beta_{xz}}{\Delta x}\left(X^{n}_{i+\frac{1}{2}\, j\,k-1} -
X^{n}_{i-\frac{1}{2}\,j\,k-1}\right).
\end{aligned}
\end{equation}
The remaining spatial derivatives $D^*_y$ and $D^*_z$ are defined
similarly with the coefficients $\alpha_y$, $\delta_y$, $\beta_{yx}$,
$\beta_{yz}$, $\alpha_z$, $\delta_z$, $\beta_{zx}$,
$\beta_{zy}$.
Note that the definition of the coefficients is identical to the one used by \cite{Lehe2013} and \cite{Nuter2014}.
The operator $D^*$ can be interpreted as a derivative ($D^*\rightarrow\partial$ for vanishing grid steps) if the coefficients fulfil the following equations~\cite{Lehe2013}:
\begin{equation}
\begin{aligned}
\alpha_x &= 1 - 2\beta_{xy} - 2\beta_{xz} - 3\delta_x\,,\\
\alpha_y &= 1 - 2\beta_{yx} - 2\beta_{yz} - 3\delta_y\,,\\
\alpha_z &= 1 - 2\beta_{zx} - 2\beta_{zy} - 3\delta_z\mathrm{.}
\end{aligned}
\label{eq:alpha}
\end{equation}

Note that at this point we have not yet made a statement about how the derivatives in \Eqsref{eq:Maxwell_Gauss_E} and \eqref{eq:Maxwell_Gauss_B} are be discretized.

\subsection{Conservation of the divergence of the fields}

Starting from the modified update equations one can find (see \ref{sec:wave_equations}), that in order to be able to obtain the discretized version of the wave equation, the Maxwell-Gauss equations must be discretized as
\begin{align}
  \label{eq:divstarb}
  \vv{D}^*\cdot \vv{B}^{n-\frac12}_{i+\frac12,j+\frac12,k+\frac12} &= 0\,,
  \\
  \label{eq:divE}
   \vv{D}\cdot \vv{E}^n_{i,j,k} &= \frac{ \rho ^n_{i,j,k}}{\varepsilon_0}\,.
\end{align}
These equations must hold at any grid point $i,j,k$.
As the necessary shifts of $\pm\frac12$ follow from the staggered grid and thus carry no actual information, they are dropped from now on.
Hence, one must use the extended stencil for the calculation of the divergence of magnetic fields and the standard stencil for the calculation of the divergence of electric fields, oppositely to the usage of discretized curl operator described in \Eqsref{eq:Maxwell_Faraday} and \eqref{eq:Maxwell_Ampere}.

It has been already stated in the literature \cite{Cowan2013} that \Eqref{eq:divstarb} is conserved automatically, if it is fulfilled in the initial moment. The same holds for \Eqref{eq:divE}, if one uses the charge conserving current deposition scheme. If both \Eqsref{eq:divstarb} and \eqref{eq:divE} are fulfilled in the initial setup, one does not need to consider them throughout the simulation.

\subsection{Initial conditions for divergence equations}
Though \Eqref{eq:divstarb} is conserved throughout the simulation, one has to be sure it is zero initially. Solving \Eqref{eq:divstarb} for the initial condition might be more complicated than solving the usual $\vv{D}\cdot\vv{B}=0$, especially if one further extends the stencil to take into account even more layers.
It is, however, possible to choose the coefficients for the extended stencil in such a way that $\vv{D}\cdot\vv{B}=0$ implies \Eqref{eq:divstarb}, making it sufficient to solve the simpler $\vv{D}\cdot\vv{B}=0$ for the initial condition.

Demanding $D_t \vv{D}\cdot \vv{B} = 0$ we find
\begin{align*}
 0 &=- \frac{1}{c} \vec{D}\cdot\partial_t\vec{B} = \vec{D}\cdot\left( \vec{D}^*\times \vec{E} \right)
 \\
  &= (D_yD_z^*-D_zD_y^*) E_x +  (D_zD_x^*-D_xD_z^*) E_y + (D_xD_y^*-D_yD_x^*) E_z\,.
\end{align*}

This imposes further conditions for the coefficients:
\begin{align}
 \beta_{ij} = \delta_j,\quad i,j=x,y,z\,.
 \label{eq:beta}
\end{align}
A scheme using a stencil that fulfils \Eqref{eq:beta} will conserve both $\vv{D}\cdot\vv{B}=0$ and \Eqref{eq:divstarb}.
In addition it can easily be shown that if the magnetic field fulfils $\vv{D}\cdot\vv{B}=0$ at any time, it also fulfils \Eqref{eq:divstarb}.

Apart from an easier initial setup, we do not see any necessity to demand \Eqref{eq:beta}.
A lot of published simulation results rely on extended stencils that do not agree with \Eqref{eq:beta}, as summarized in \Tabref{tab:norms}.
For the time being, we will consider this constraint as optional and present results both ways.

\section{Dispersion relation}\label{sec:dispersion_relation}
The dispersion relation for the discretized Maxwell's equations as described by \Eqsref{eq:Maxwell_Faraday} and \eqref{eq:Maxwell_Ampere} is given by
\begin{gather}
s_\omega^2
=s_x^2 A_x  + s_y^2 A_y + s_z^2 A_z \label{eq:dispersion_relation}
\end{gather}
with the abbreviations
\begin{equation}
\begin{aligned}
s_\omega &= \frac{\sin\left(\frac{1}{2}\omega \Delta t\right)}{c\Delta t}\,, \\
 s_{\{x,y,z\}} &= \frac{\sin\left(\frac{1}{2}k_{\{x,y,z\}} \Delta \{x,y,z\}\right)}{\Delta \{x,y,z\}}
\end{aligned}
\end{equation}
and
\begin{equation}
\begin{aligned}
A_x ={}&\alpha_x + 2\beta_{xy}\cos(k_y\Delta y) + 2\beta_{xz}\cos(k_z\Delta z)\\
       &+ \delta_x(1+2\cos(k_x\Delta x))\,, \\
A_y ={}& \alpha_y + 2\beta_{yx}\cos(k_x\Delta x) + 2\beta_{yz}\cos(k_z\Delta z)\\
       &+ \delta_y(1+2\cos(k_y\Delta y))\,, \\
A_z ={}& \alpha_z + 2\beta_{zx}\cos(k_x\Delta x) + 2\beta_{zy}\cos(k_y\Delta y)\\
       & + \delta_z(1+2\cos(k_z\Delta z))\mathrm{.}
\end{aligned}
\end{equation}
From this it can be shown (see  \ref{sec:betasymmetry}), that the dispersion relation does not depend on all six $\beta$ coefficients independently, but only on the symmetric part of
\begin{equation}
 \label{eqn:beta_symmetric_part}
 \hat{\beta}_{ij} \equiv \Delta x_j^2 \beta_{ij}\,.
\end{equation}
This reduces the number of free coefficients.

We need to make sure that \Eqref{eq:dispersion_relation} must remain well defined, i.\,e.,
\begin{gather}
\label{eq:constraints}
0\leq \left( c\Delta t \right)^2\left( s_x^2 A_x  + s_y^2 A_y + s_z^2 A_z \right) \leq 1\,, \quad 0\leq k_i\leq \pi\mathrm{.}
\end{gather}
The right hand inequality is equivalent to the Courant-Friedrichs-Lewy (CFL) condition, while equality corresponds to a conditionally stable scenario.
Any

The stencil reduces to the well-known Yee stencil for vanishing $\beta$ and
$\delta$. In this case all $\alpha$ coefficients must be unity as enforced by \Eqref{eq:alpha}, and the dispersion relation simplifies to the well-known Yee dispersion~\cite{Birdsall:1991}. Note, that in one dimensional case the Yee scheme is dispersion-free for $c\Delta t=\Delta x$. However, as this time step is only conditionally stable, the introduction of particles will turn the PIC code unstable. Reducing the time step generically introduces dispersive effects.

Following this observation we want to emphasize that the time step is an important parameter that must also be taken into account while optimizing the stencil.
This was already recognized in \Refcite{Cowan2013}, where the authors proposed to find a time step numerically such that the correct group velocity of the laser in a plasma with a given density is ensured. In the following section we will introduce a norm to evaluate the stencils depending on the coefficients as well as the time step.

\section{A norm function for dispersion relation}\label{sec:norm}
In order to quantify the search for the coefficients we propose a norm for
dispersion relation
\begin{gather}
\label{eq:norm}
f_w[\omega] = \Delta x\Delta y\Delta z\int_{0}^{\frac{\pi}{\Delta x}}\dd{k_x} \int_{0}^{\frac{\pi}{\Delta y}}\dd{k_y} \int_{0}^{\frac{\pi}{\Delta z}}\dd{k_z} w(\vv{k})\cdot\left( \omega(\vv{k}) - c |\vv k| \right)^2\mathrm{.}
\end{gather}
The norm introduced in \Eqref{eq:norm} is the variance, a measure for the distance of the grid dispersion relation $\omega$, as given by \Eqref{eq:dispersion_relation}, to the free-space dispersion relation $\omega_0=c|\vv{k}|$.
As the grid dispersion relation in turn depends on the coefficients of the stencil $\alpha_i, \beta_{ij}, \delta_i$ and the time step $\Delta t$, we can now optimize these with respect to the function $f[\omega]$.
The norm defined in \Eqref{eq:norm} allows the choice of a weight function $w=w(k_x, k_y, k_z)$ to optimize the stencil with respect to special physical setups, for example, along a given direction by taking a non-vanishing $w$ in a given cone, or introducing limits on $|\vec{k}|$ to select regions of interest in frequency space.

The optimal stencil depends on a number of choices, e.\,g.,
\begin{itemize}
 \item the number of dimensions,
 \item the weight function $w(\vv{k})$,
 \item the grid aspect ratios $Y=\frac{\Delta y}{\Delta x}$ and $Z=\frac{\Delta z}{\Delta x}$,
 \item choosing to demand any number of further identities, e.\,g., $\delta_i=\delta_j$,
 \item the choice whether to adhere to \Eqref{eq:beta}\,,
\end{itemize}
while in any case \Eqref{eq:alpha} must be fulfilled.
Depending on these choices the number of free parameters varies between 1 and 7.

In order to minimize \Eqref{eq:norm} with respect to the coefficients, we employ constrained optimization using sequential least squares programming \cite{Kraft1988,SciPy}.
The constraints are given from \Eqref{eq:constraints}, the target function by \Eqref{eq:norm}.
Additionally a multidimensional cuboid is chosen in parameter space to act as the search space.
The bounds of these cuboid are added to the optimization as additional constraints, which prevents the optimizer from running away towards non-sensical solutions.
Of course this cuboid must be chosen large enough such that it encloses any plausible stencil.
The optimization is started on a grid of possible initial points within the search space, in order to increase the chances of finding the global minimum. Our code for optimizing the stencils can be found in \Refcite{optimization_code}.

\section{Optimized stencils}\label{sec:optimized_stencils}
\begin{table}[tb]
\centering
\caption{
Norm and coefficients for different schemes, available in the literature, and results found through our minimization algorithm in 2D with  $\Delta x = \Delta y$.
In this case Pukhov's scheme is a special case of Cowan's scheme.
The minimized results which fulfil \Eqref{eq:beta} are computed with fixed $c\frac{\Delta t}{\Delta x}$.}
\label{tab:norms}
\begin{tabular}{cccccccccc}
\toprule
Scheme& $f_{w_0}(\omega)$ & $c\frac{\Delta t}{\Delta x}$ &
$\beta_{xy}$ = $\beta_{yx}$   & $\delta_x$ & $\delta_y$ & \!\Eqref{eq:beta}\!
\\\midrule
\textsc{Yee} \cite{Yee1966}   & $1.06$ & $\frac{0.95}{\sqrt{2}}$ & 0 & 0 & 0 & \cmark
\\
NDFX \cite{Pukhov1999}        & $0.83$ & $1.00$                   & $0.125$ & 0 & 0 & \xmark
\\
Cowan \cite{Cowan2013}        & $0.93$ & $0.999$                  & $0.125$ & 0 & 0 & \xmark
\\
\textsc{Lehe} \cite{Lehe2013} & $1.04$ & $0.96$                   & $0.125$ & $-0.021$ & 0& \xmark
\\
min.~1                        & $0.08$ & $0.686$                  & $0.110$ & $-0.125$ & $-0.125$& \xmark
\\
min.~2                        & $0.63$ & $\frac{0.95}{\sqrt{2}}$  & $-0.013$ & $-0.013$ & $-0.013$& \cmark
\\
min.~3                        & $0.42$ & $0.500$                  & $-0.065$ & $-0.065$ & $-0.065$& \cmark
\\
min.~4                        & $0.36$ & $0.100$                  & $-0.125$ & $-0.125$ & $-0.125$& \cmark
\\\bottomrule
\end{tabular}
\end{table}

A number of numerically optimized stencils in 2D with  $\Delta x = \Delta y$ are shown in \Tabref{tab:norms} and compared to published sets of coefficients.
In the beginning we chose a unity weighting function $w_0(\vv{k})=1$ and fixed $\delta_x=\delta_y$ yielding manifestly symmetric stencils.
We also fixed $\beta_{xy}=\beta_{yx}$, which does not impose any actual restriction for the stencil but cancels out a spurious degree of freedom, because it only depends on the symmetric part \Eqref{eqn:beta_symmetric_part}.
The decreases the number of free parameters to 3 in case of min.~1.
In the cases min.~2 to min.~4, where we additionally chose to obey \Eqref{eq:beta}, only 2 free parameters remain.
In these cases the minimization tends towards arbitrarily short time steps and it is necessary to give a lower bound to the time step.
The minimization algorithm generically finds time steps which are below the CFL-condition and should be stable also in the presence of particles.

Yee's scheme is said to be stable in the presence of particles with a widely used time step of $95\%$ of the CFL-condition.
Pukhov states that the NDFX scheme is stable even with a time step $c\Delta t=\Delta x$, assuming $\Delta x\leq\min(\Delta y,\Delta z)$ \cite{Pukhov1999}.
According to our norm the NDFX scheme is already a significant improvement over Yee's scheme.
In 2D, Cowan's scheme differs from Pukhov's scheme only in that it chooses a slightly smaller time step which will improve the group velocity and dispersion around the laser frequency in the plasma \cite{Cowan2013} while making the overall norm slightly worse\footnote{In 3D Cowans scheme uses an additional set of grid points which are not included neither in our definition nor in the NDFX.}.
Lehe's scheme is based upon choosing a time step first.
When $c\Delta t=\Delta x$ is chosen it also reduces to Pukhov's scheme \cite{Lehe2013}, Lehe proposes a choice of $c\Delta t=0.96 \Delta x$.
The norm of Lehe's scheme is almost as high as the norm of Yee's scheme.
In order to understand that we have to look at the resulting dispersion relations.

\begin{figure}[tbp]
  \centering
  \includegraphics[scale=0.75] {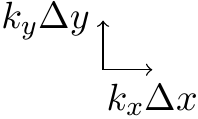}
  \hspace{-0.8cm}
  \includegraphics[scale=0.75] {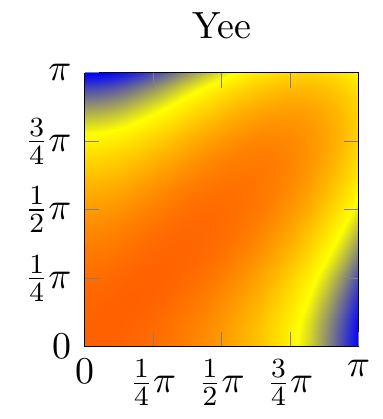}
  \hspace{-0.4cm}
  \includegraphics[scale=0.75] {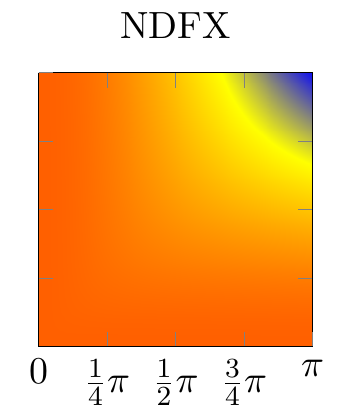}
  \hspace{-0.4cm}
  \includegraphics[scale=0.75] {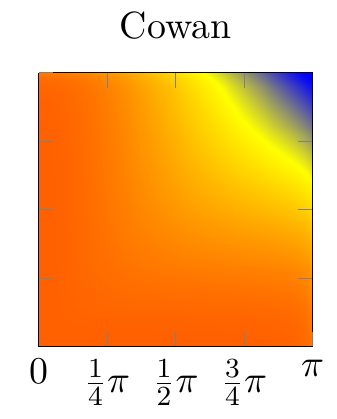}
  \hspace{-0.4cm}
  \includegraphics[scale=0.75] {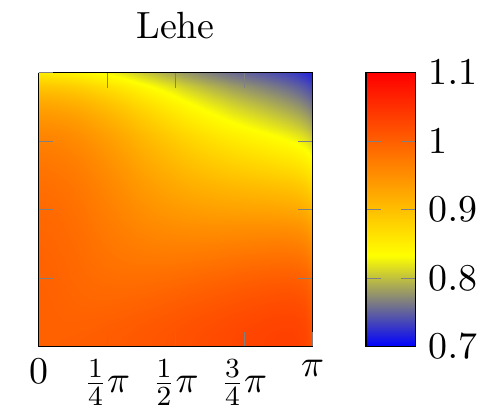}
  \\
  \includegraphics[scale=0.75] {plot.pdf}
  \hspace{-0.8cm}
  \includegraphics[scale=0.75] {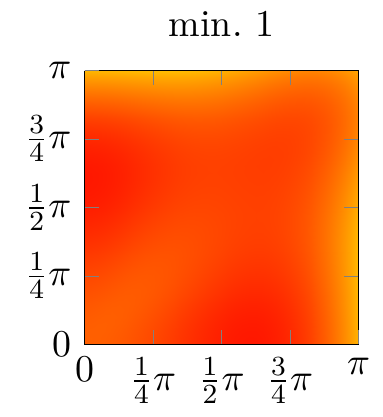}
  \hspace{-0.4cm}
  \includegraphics[scale=0.75] {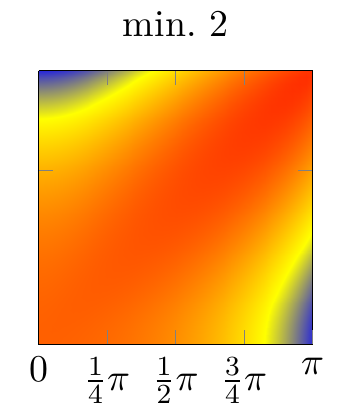}
  \hspace{-0.4cm}
  \includegraphics[scale=0.75] {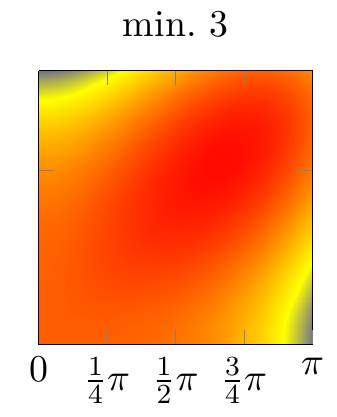}
  \hspace{-0.4cm}
  \includegraphics[scale=0.75] {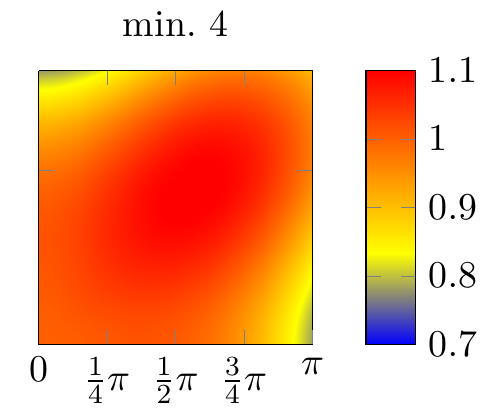}
  \caption{
    The color encodes the phase velocities $\frac{v_{\text{ph}}}{c}$ for the minimized stencils listed in \Tabref{tab:norms}.
    The corresponding minima and maxima are listed in \Tabref{tab:phase_velocity}.
    It can be seen, that our proposed scheme (lower right box) has the minimal deviations of the phase velocity from the speed of light.
     }
  \label{fig:phase_velocities}
\end{figure}
\begin{table}[tb]
\caption{Maximal and minimal phase velocities for the setups listed in \Tabref{tab:norms} and shown in \Figref{fig:phase_velocities}.}
\label{tab:phase_velocity}
\hspace*{\fill}
\begin{tabular}[t]{ccc}
\toprule
Scheme & $\min(\frac{v_{\text{ph}}}{c})$ &$\max(\frac{v_{\text{ph}}}{c})$  \\\midrule
\textsc{Yee} & $0.70$ & $1$             \\
NDFX & $0.71$ & $1$          \\
\textsc{Lehe} & $0.72$ & $1.03$         \\
\bottomrule
\end{tabular}
\hspace*{\fill}
\begin{tabular}[t]{ccc}
\toprule
 Scheme & $\min(\frac{v_{\text{ph}}}{c})$ &$\max(\frac{v_{\text{ph}}}{c})$  \\\midrule
min.~1 & $0.90$ & $1.08$ \\
 min.~2 & $0.72$ & $1.05$ \\
 min.~3 & $0.76$ & $1.09$ \\
 min.~4 & $0.78$ & $1.11$ \\
\bottomrule
\end{tabular}
\hspace*{\fill}
\end{table}
\Figref{fig:phase_velocities} sketches the phase velocity on the grid, defined as
\begin{gather}
v_{\text{ph}} = \frac{\omega}{|\vv k|}\,,
\end{gather}
and \Tabref{tab:phase_velocity} gives an overview about the maximal deviations of the phase velocity from the speed of light.
The plots clearly show, that for the proposed sets of coefficients the phase velocity has less deviations from $c$ than in any of the other sets available.
As the norms do not vanish, the phase velocity on the grid is not equal to the speed of light for all $\vec k$.
Lehe's set of coefficients and our proposed sets have phase velocities larger than the speed of light $c$.
In the case of the minimized scheme this is due to our choice of optimization metric.
If, initially, the phase velocity is too small in one region, the stencil will be modified to increase it.
This in turn causes the phase velocity to overshoot in other regions.
As the norm weights all regions equally, an equilibrium between too small and too large phase velocities across the whole $k$ space is found.
This is also the reason why the norm of Lehe's scheme is so high: it was optimized for propagation along the $x$ axis, while our norm optimizes for the whole $\vv k$ space.
This speedup of the phase velocity in the minimized cases is smaller than the slowdown of phase velocities in the Yee, Pukhov and Cowan setups.

\begin{figure}[tbp]
  \centering
  \includegraphics[scale=0.75] {plot.pdf}
  \hspace{-0.8cm}
  \includegraphics[scale=0.75] {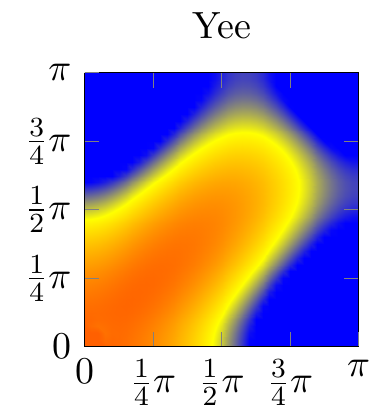}
  \hspace{-0.4cm}
  \includegraphics[scale=0.75] {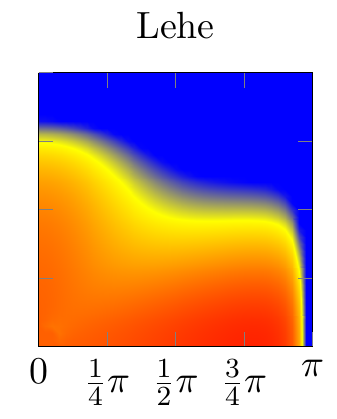}
  \hspace{-0.4cm}
  \includegraphics[scale=0.75] {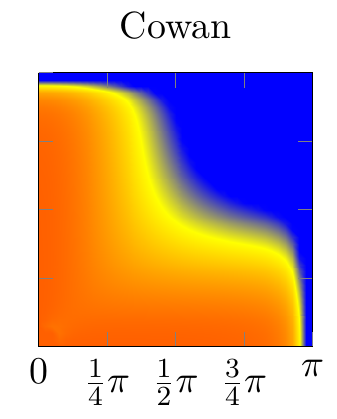}
  \hspace{-0.4cm}
  \includegraphics[scale=0.75] {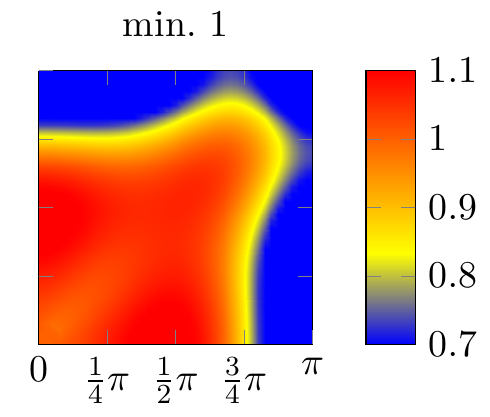}%
  \caption{
    The color encodes the group velocities $\frac{|v_{\mathrm{g}}|}{c}$ for some of the stencils listed in \Tabref{tab:norms}.
  }
  \label{fig:group_velocities}
\end{figure}
As can be seen in \Figref{fig:group_velocities}, the group velocities
\[
 |v_{\mathrm{g}}|=\left|\frac{\dd{\omega}}{\dd{\vec{k}}}\right|
\]
show the same qualitative behaviour as the phase velocities.
In general, taking the derivative amplifies the deviations from $c$.
Please note that the modified schemes, even when optimized with respect to the phase velocity, generally have group velocities closer to $c$ and will model the laser pulse propagation more accurately than Yee's scheme.
As is the case with the phase velocities in both Lehe's scheme and in our minimized schemes the peak group velocity is slightly faster than $c$.

\subsection{Reproducing the published stencils using optimization}

Interestingly, if certain constraints and conditions are employed in the optimization, it is, in 2D with $\Delta x=\Delta y$, possible to reproduce the schemes by Pukhov and Lehe.
If the lower bound for the time step is set to $\Delta t = \frac{\Delta x}{c}$ the optimization produces the NDFX scheme.
Setting the bounds for $\delta_y$ tightly around zero and the upper bound for the time step to $\Delta t = 0.96  \frac{\Delta x}{c}$ reproduces Lehe's scheme with $\delta_x=-0.021$.

\subsection{Optimizing close to a specific axis of propagation}

Choosing a weight function, e.\,g.
\begin{equation}
 \label{eqn:norm_x_axis}
 w = e^{-\left( \frac{k_y\Delta y}{0.1} \right)^2}\,,
\end{equation}
enables us to find a stencil which has good dispersion properties when propagating along the $x$ axis, i.\,e. with small $k_y\Delta y$.
Optimizing numerically with respect to this norm we find a stencil that is, on the axis, better than Lehe's stencil, see \Tabref{tab:norms_weighted}.
With a time step of $c\frac{\Delta t}{\Delta x}=1.0$ we could not find a stencil that is better than NDFX on the axis,
while a time step of $c\frac{\Delta t}{\Delta x}=0.999$ allowed for an improvement of Cowan's scheme. It is also interesting to note, that our approach allows to find coefficients such that the scheme is stable even if the time step $c\Delta t > \Delta x$.
\begin{table}[htb]
\centering
\caption{
Norm with weight function from \Eqref{eqn:norm_x_axis} in 2D with  $\Delta x = \Delta y$ and coefficients for the schemes referenced in text and two optimized schemes found through our minimization algorithm with fixed $c\frac{\Delta t}{\Delta x}$.}
\label{tab:norms_weighted}
\begin{tabular}{ccccccccc}
\toprule
Scheme& $f_{w}(\omega)$ & $c\frac{\Delta t}{\Delta x}$ & $\beta_{xy}$ = $\beta_{yx}$   & $\delta_x$ & $\delta_y$
\\\midrule
\textsc{Yee} \cite{Yee1966}   & $0.036$   & $\frac{0.95}{\sqrt{2}}$ & 0       & 0        & 0
\\
\textsc{Lehe} \cite{Lehe2013} & $0.008$   & $0.96$                  & $0.125$ & $-0.021$ & $0.000$
\\
min.~5                     & $0.002$   & $0.96$                  & $0.133$ & $-0.017$ & $-0.019$
\\
NDFX \cite{Pukhov1999}        & $1.5\cdot10^{-7}$ & $1.00$                  & $0.125$ & 0        & 0
\\
Cowan \cite{Cowan2013}        & $4.5\cdot10^{-5}$ & $0.999$                  & $0.125$ & 0        & 0
\\
min.~6                     & $6.2\cdot10^{-7}$ & $0.999$                  & $0.128$ & $-0.0005$ & 0
\\\bottomrule
\end{tabular}
\end{table}

\subsection{Non-square grid aspect ratio}

\begin{figure}[tb]
  \includegraphics[scale=0.75] {plot.pdf}
  \hspace{-0.8cm}
  \includegraphics[scale=0.75] {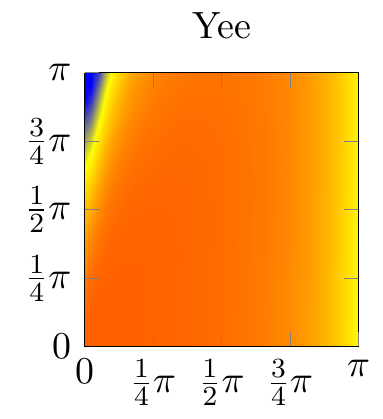}
  \hspace{-0.4cm}
  \includegraphics[scale=0.75] {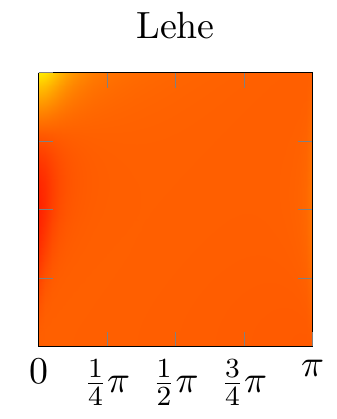}
  \hspace{-0.4cm}
  \includegraphics[scale=0.75] {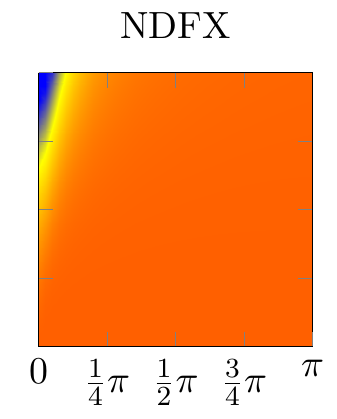}
  \hspace{-0.4cm}
  \includegraphics[scale=0.75] {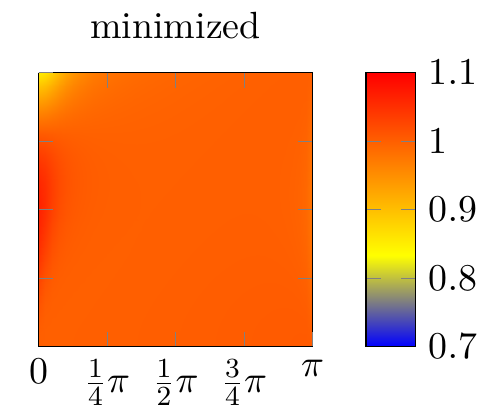}
  \caption{
    The color encodes the phase velocities $\frac{v_{\text{ph}}}{c}$ for setups with $\Delta y = 10 \Delta x$.
    The norms $|\omega_G|$ of these dispersion relation are from left to right $20.7$, $2.1$, $0.24$, $0.059$.
    Of these stencils only the Yee stencil adheres to \Eqref{eq:beta}.
    Note that \Figsref{fig:phase_velocities} and \ref{fig:phase_velocities_2} use the same colorbar.
  }
  \label{fig:phase_velocities_2}
\end{figure}

An advantage of our scheme is, that due to the minimization algorithm, an optimal set of coefficients for any ratio $\Delta x / \Delta y$ can easily be constructed, while analytically prescribed coefficients have to make tradeoffs.
\Figref{fig:phase_velocities_2} shows the phase velocities for a grid with a large aspect ratio $\Delta y = 10\Delta x$.
Similar to \Figref{fig:phase_velocities} we find that the dispersion relation of the proposed scheme is closer to the free dispersion relation than any other scheme available.


\section{Numerical examples}\label{sec:numerical}
 \begin{figure}[tb]
  \includegraphics[width=1 \textwidth] {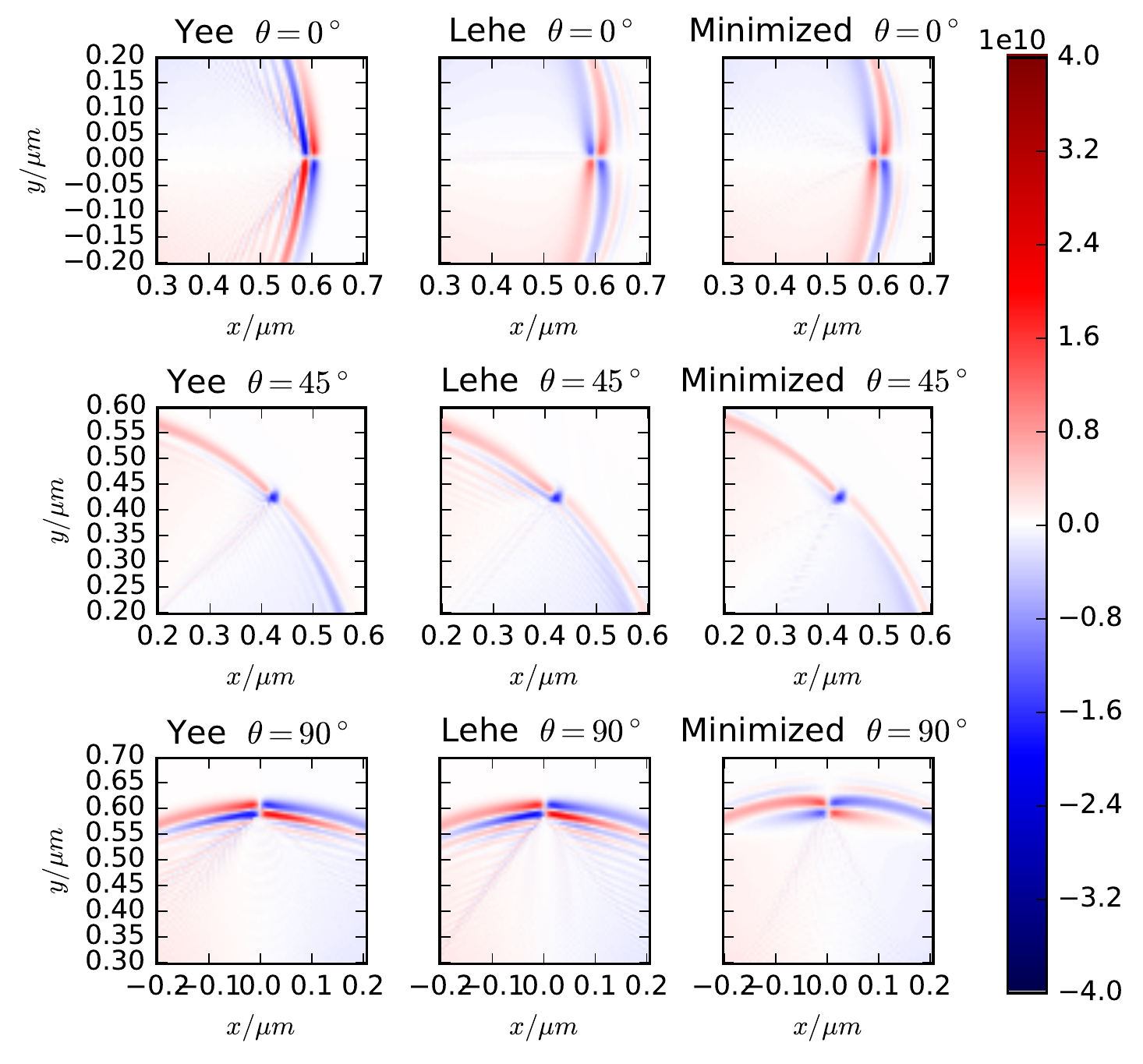}
  \caption{
    The transverse field strength in arbitrary units is plotted to visualize the numerical Cherenkov radiation of an electron bunch with $\gamma=10$ for various solvers and different angles $\theta$ with respect to the $x$-axis.
    The angles are from top to bottom $\theta=0^\circ$, $\theta=45^\circ$, $\theta=90^\circ$, the solvers from left to right are Yee, Lehe, the proposed solver ``min.~1'' (see \Tabref{tab:norms}).
  }
  \label{fig:cherenkov_angular}
\end{figure}

\Figref{fig:cherenkov_angular} shows results of a PIC simulation
performed with the EPOCH code~\cite{ArberEPOCH} for different Maxwell solvers.
All Maxwell solvers presented in this paper, including the proposed scheme, are submitted to be included into the EPOCH code.
The plots show the numerical Cherenkov radiation for an relativistic electron bunch with $\gamma=10$ propagating under different angles $\theta$ with respect to the $x$-axis. Along the $x$-axis, $\theta=0^\circ$, Lehe's solver produces only a negligible amount of
artificial Cherenkov radiation but draws a tail behind the electron bunch (see
also \cite{Lehe2013}), which is also due to the numerical dispersion of the grid. Our proposed solver emits significantly less numerical
Cherenkov radiation compared to the Yee case and does not generate the tail
structure.  For electron bunches propagating under the angle to the $x$ axis, our scheme always generates
significantly less numerical Cherenkov radiation. This is useful in scenarios when several electron bunches propagating under different angles have to be modelled, for example, in astrophysical cases or when modelling multi-laser-beam irradiation of targets.

\section{Conclusions}\label{sec:conclusions}
A scheme to minimize the phase velocity errors in FDTD schemes using extended stencils to calculate the spatial derivative was proposed. By introducing the norm function, describing the difference of the grid dispersion relation to the free-space dispersion relation, one can generate an optimal set of coefficients for a given physical problem, e.g. to get rid of the numerical dispersion along a given axis or directions.
Our scheme generates significantly less numerical Cherenkov radiation than Yee's scheme and generates no tail as compared to Lehe's scheme. Although the proposed scheme (or any of the FDTD schemes) does not completely remove the Cherenkov instability, it can significantly decrease it. In principle, it is also possible to apply our idea for minimization of the numerical group velocity error or for minimization of the numerical error of both phase and group velocities at the same time. Another advantage of our scheme is that an optimal set of parameters can be constructed for any ratio of grid steps. In general, our scheme provides a more homogeneous and close to $c$ distribution of phase velocity for all directions, which makes this scheme suitable for multi-beam irradiation scenarios and scenarios relevant to astrophysics.

\section*{Acknowledgments}
This work was supported by the Helmholtz Association (Young Investigator's Group
VH-NG-1037). The authors gratefully acknowledge the usage of the open-source
visualization software Matplotlib \cite{Hunter2007} and IPython
\cite{Perez2007}.

\appendix
\section{Wave equations emerging from the Leapfrog scheme}
\label{sec:wave_equations}

In this appendix we show how the update equations of the Maxwell solver imply wave equations for the fields.
This computation is a necessary homework that gives us additional information on how the update equations need to be used.
Note that actual PIC codes might use half steps in order to calculate both $\vv{E}$ and $\vv{B}$ at the same point in time.
In those cases the update equations in the code are different from the ones presented here.
In case of the EPOCH code it can be shown easily that they are equivalent \cite{ArberEPOCH}.

\subsection{Numerical Scheme}
In order to design a consistent numerical scheme, we need to make sure that the numeric forms of Maxwell's equations and those derived from them are equivalent to their continuum counterparts.
We start with the update equations, which are nothing but some choice of a discretization of \Eqsref{eq:Maxwell_Faraday} and \eqref{eq:Maxwell_Ampere}.
We then find the corresponding numeric forms of $\vv{\nabla}\cdot \vv{E} = \frac{\rho}{\varepsilon_0}$,
$\vv{\nabla}\cdot \vv{B} = 0$
and $0 = \partial_t \rho+\vv{\nabla}\cdot \vv{J}$
which we need to ensure, such that we find also the correct numerical forms of the wave equations
\begin{align}
 \label{eq:wave_equation_magnetic_current}
 \left( \frac{1}{c^2}\partial_t^2 - \Delta \right) \vv{B} &=  \mu_0 \vv{\nabla}\times \vv{J}\,,
 \\
 \label{eq:wave_equation_electric_current}
 \left( \frac{1}{c^2}\partial_t^2 - \Delta \right) \vv{E} &=   - \frac{1}{\varepsilon_0}\vv{\nabla}\rho- \mu_0\partial_t\vv{J}\,.
\end{align}

The update equations
\begin{align}
  \label{eq:update1}
  B^{n+\frac12}_{x|i,j+\frac12,k+\frac12} &= B^{n-\frac12}_{x|i,j+\frac12,k+\frac12} + \Delta t \left( D^*_z E^n_{y|i,j+\frac12,k+\frac12} - D^*_y E^n_{z|i,j+\frac12,k+\frac12} \right)\,, \\
  B^{n+\frac12}_{y|i+\frac12,j,k+\frac12} &= B^{n-\frac12}_{y|i+\frac12,j,k+\frac12} + \Delta t \left( D^*_x E^n_{z|i+\frac12,j,k+\frac12} - D^*_z E^n_{x|i+\frac12,j,k+\frac12} \right)\,, \\
  B^{n+\frac12}_{z|i+\frac12,j+\frac12,k} &= B^{n-\frac12}_{z|i+\frac12,j+\frac12,k} + \Delta t \left( D^*_y E^n_{x|i+\frac12,j+\frac12,k} - D^*_x E^n_{y|i+\frac12,j+\frac12,k} \right)\,, \\
  E^{n+1}_{x|i+\frac12,j,k} &= E^{n}_{x|i+\frac12,j,k} \!+\! c^2\Delta t \left( D_y B^{n+\frac12}_{z|i+\frac12,j,k} - D_z B^{n+\frac12}_{y|i+\frac12,j,k}-  \mu_0 J^{n+\frac12}_{x|i+\frac12,j,k} \right) , \\
  \label{eq:update5}
  E^{n+1}_{y|i,j+\frac12,k} &= E^{n}_{y|i,j+\frac12,k} \!+\! c^2\Delta t \left( D_z B^{n+\frac12}_{x|i,j+\frac12,k} - D_x B^{n+\frac12}_{z|i,j+\frac12,k}- \mu_0 J^{n+\frac12}_{y|i,j+\frac12,k}  \right) ,\\
  \label{eq:update6}
  E^{n+1}_{z|i,j,k+\frac12} &= E^{n}_{z|i,j,k+\frac12} \!+\! c^2\Delta t \left( D_x B^{n+\frac12}_{y|i,j,k+\frac12} - D_y B^{n+\frac12}_{x|i,j,k+\frac12}-  \mu_0 J^{n+\frac12}_{z|i,j,k+\frac12} \right)
\end{align}
result from replacing the second order accurate, central stencils $D_i$ in the update equations for the magnetic field of Yee's scheme by the extended stencils $D_i^*$.
The update equations can be written in vectorial form as
\begin{align}
 \label{eq:maxwell_faraday_extdnum_charged}
 D_t \vv{B}^n_{i+\frac12,j+\frac12,k+\frac12} &= -\vv{D}^*\times \vv{E}^n_{i+\frac12,j+\frac12,k+\frac12}\,,
 \\
 \label{eq:maxwell_ampere_extdnum_charged}
  \frac{1}{c^2}D_t \vv{E}^{n+\frac12}_{i,j,k}  &=\vv{D}\times \vv{B}^{n+\frac12}_{i,j,k} -\mu_0\vv{J}^{n+\frac12}_{i,j,k}\,,
\end{align}
where the components correspond to the update equations when shifted by one half step into the corresponding direction.
When we apply the numerical divergence operator $\vv{D}$ to \Eqref{eq:maxwell_ampere_extdnum_charged}, this shifting is automatically taken care of by the stencil in the numerical divergence operator.
Suppressing the indices for clarity, we obtain
\begin{align}
 0 &= \vv{D}\cdot \left( \vv{D}\times \vv{B} \right)
 \\
 &= \vv{D}\cdot \left( \frac{1}{c^2}D_t \vv{E} +  \mu_0\vv{J} \right)
 \\
 &= \mu_0\varepsilon_0 D_t \vv{D}\cdot \vv{E} +  \mu_0\vv{D}\cdot\vv{J}\,.
\end{align}
When we define
\begin{align}
 \vv{D}\cdot \vv{E} =: \frac{\rho_E}{\varepsilon_0}
\end{align}
we can deduce
\begin{align}
 \label{eq:numerical_continuity_equation}
 0 &=  D_t \rho_E +  \vv{D}\cdot\vv{J}\,.
\end{align}
With all indices in place, this actually means
\begin{align}
 0 {}={}&  D_t (\rho_E)^{n+\frac12}_{i,j,k} +  \vv{D}\cdot\vv{J}^{n+\frac12}_{i,j,k}
 \\
 {}={}& \frac{1}{\Delta t} \left( (\rho_E)^{n+1}_{i,j,k} - (\rho_E)^{n}_{i,j,k} \right) + \frac{1}{\Delta x} \left( J^{n+\frac12}_{x|i+\frac12,j,k} - J^{n+\frac12}_{x|i-\frac12,j,k}\right) \\
 \nonumber
 & +\frac{1}{\Delta y} \left( J^{n+\frac12}_{y|i,j+\frac12,k} - J^{n+\frac12}_{y|i,j-\frac12,k} \right) + \frac{1}{\Delta z} \left( J^{n+\frac12}_{z|i,j,k+\frac12} - J^{n+\frac12}_{z|i,j,k-\frac12} \right) \,.
\end{align}
The index $E$ indicates that this charge density $\rho_E$ is at this point not an external input but a function of the electric field which is in turn a function of the currents at all earlier times and the charge density in the initial condition.
If we have both $\rho$ and $\vv{J}$ as external inputs, we need to make sure that $\rho$ and $\rho_E$ are consistent.
This is the task of the current weighting scheme, which calculates the $\vv{J}^{n+\frac12}_{y|i,j,k}$ such that $\rho^n = \rho^n_E \Rightarrow \rho^{n+1} = \rho^{n+1}_E$ \cite{Esirkepov2001}.
At this point it becomes clear that this would become more complicated if we used the extended stencil also in \Eqref{eq:maxwell_ampere_extdnum_charged}.

Taking the extended numerical divergence of \Eqref{eq:maxwell_faraday_extdnum_charged}, we find
\begin{align}
 0 &= -\vv{D}^*\cdot \left( \vv{D}^*\times \vv{E} \right)
 \\
 &= \vv{D}^*\cdot \left( D_t \vv{B}^n_{i+\frac12,j+\frac12,k+\frac12} \right)
 \\
 &= D_t \left( \vv{D}^*\cdot \vv{B}^n_{i+\frac12,j+\frac12,k+\frac12} \right)\,.
\end{align}

\subsection{Magnetic Wave Equation}
Let us have a look at the wave equations.
Note, that we can shift \Eqref{eq:update1} by one timestep to obtain
\begin{align}
 \label{eq:update1_shifted}
 -B^{n-\frac32}_{x|i,j+\frac12,k+\frac12} &= - B^{n-\frac12}_{x|i,j+\frac12,k+\frac12} + \Delta t \left( D^*_z E^{n-1}_{y|i,j+\frac12,k+\frac12} - D^*_y E^{n-1}_{z|i,j+\frac12,k+\frac12} \right)\,.
\end{align}
We can then start from \Eqref{eq:update1} and insert \Eqsref{eq:update5}, \eqref{eq:update6} and \eqref{eq:update1_shifted} to obtain
\begin{align}
\nonumber
  B^{n+\frac12}_{x|i,j+\frac12,k+\frac12} ={}& B^{n-\frac12}_{x|i,j+\frac12,k+\frac12} + \Delta t \left( D^*_z E^n_{y|i,j+\frac12,k+\frac12} - D^*_y E^n_{z|i,j+\frac12,k+\frac12} \right) \\
  \nonumber ={}& 
      B^{n-\frac12}_{x|i,j+\frac12,k+\frac12} + \Delta t \left( D^*_z E^{n-1}_{y|i,j+\frac12,k+\frac12} -   D^*_y E^{n-1}_{z|i,j+\frac12,k+\frac12}  \right) \\
      \nonumber &+ c^2\Delta t^2 D^*_z \left( D_z B^{n-\frac12}_{x|i,j+\frac12,k+\frac12} - D_x B^{n-\frac12}_{z|i,j+\frac12,k+\frac12}  \right)
      \\
      \nonumber &- c^2\Delta t ^2  D^*_y \left( D_x B^{n-\frac12}_{y|i,j+\frac12,k+\frac12} - D_y B^{n-\frac12}_{x|i,j+\frac12,k+\frac12}\right)
      \\\nonumber & -c^2\Delta t^2 \left(  D^*_z J^{n-\frac12}_{y|i,j+\frac12,k+\frac12} +  D^*_y J^{n-\frac12}_{z|i,j+\frac12,k+\frac12} \right)
  \\
  \nonumber={}& 
      2B^{n-\frac12}_{x|i,j+\frac12,k+\frac12} -B^{n-\frac32}_{x|i,j+\frac12,k+\frac12} \\
      \nonumber &+ c^2\Delta t^2 D^*_z \left( D_z B^{n-\frac12}_{x|i,j+\frac12,k+\frac12} - D_x B^{n-\frac12}_{z|i,j+\frac12,k+\frac12}  \right)
      \\
      \nonumber &- c^2\Delta t ^2  D^*_y \left( D_x B^{n-\frac12}_{y|i,j+\frac12,k+\frac12} - D_y B^{n-\frac12}_{x|i,j+\frac12,k+\frac12}\right)
      \\\nonumber &+ c^2\Delta t^2 D^*_x \left( D_x B^{n-\frac12}_{x|i,j+\frac12,k+\frac12} - D_x B^{n-\frac12}_{x|i,j+\frac12,k+\frac12}  \right)
      \\\nonumber & -c^2\Delta t^2 \left(  D^*_z J^{n-\frac12}_{y|i,j+\frac12,k+\frac12} +  D^*_y J^{n-\frac12}_{z|i,j+\frac12,k+\frac12} \right)\,.
\end{align}
Recognizing the second order temporal derivative stencil $D_t^2$ we find
\begin{align}
   \frac{1}{c^2} D_t^2 B^{n-\frac12}_{x|i,j+\frac12,k+\frac12}={}& \begin{multlined}[t]
\left( \vv{D}^*\cdot\vv{D} \right) B^{n-\frac12}_{x|i,j+\frac12,k+\frac12}- D_x \left( \vv{D}^*\cdot \vv{B}^{n-\frac12}_{i,j+\frac12,k+\frac12}\right)
   \\\nonumber
       - D^*_z \mu_0 J^{n-\frac12}_{y|i,j+\frac12,k+\frac12}
       +  D^*_y \mu_0 J^{n-\frac12}_{z|i,j+\frac12,k+\frac12}\,,
                                              \end{multlined}
   \intertext{where we also used $D_iD_j^*-D_j^*D_i=0$.
   When we accept $ \vv{D}^*\cdot\vv{D} =  \vv{D}\cdot\vv{D}^*$ as our new extended Laplacian operator $\Delta^*$, we find}
   \frac{1}{c^2} D_t^2 B^{n-\frac12}_{x|i,j+\frac12,k+\frac12}
    ={}& \Delta^* B^{n-\frac12}_{x|i,j+\frac12,k+\frac12} + \mu_0\left( \vv{D}^*\times\vv{J}^{n-\frac12}_{i,j+\frac12,k+\frac12} \right)_x\,,
\end{align}
under the condition that
\begin{align}
 \vv{D}^*\cdot \vv{B}^{n-\frac12}_{i+\frac12,j+\frac12,k+\frac12} &= 0
\end{align}
at any point.
At this point we succeeded in finding the correct numerical form of \Eqref{eq:wave_equation_magnetic_current}
\begin{align}
 \frac{1}{c^2} D_t^2 \vv{B}^{n-\frac12}_{i+\frac12,j+\frac12,k+\frac12} = \Delta^* \vv{B}^{n-\frac12}_{i+\frac12,j+\frac12,k+\frac12} + \mu_0 \vv{D}^*\times\vv{J}^{n-\frac12}_{i+\frac12,j+\frac12,k+\frac12} \,.
\end{align}

\subsection{Electric Wave Equation}

For the electric field we find
\begin{align}
 E^{n+1}_{x|i+\frac12,j,k} ={}& E^{n}_{x|i+\frac12,j,k} + c^2\Delta t \left( D_y B^{n+\frac12}_{z|i+\frac12,j,k} - D_z B^{n+\frac12}_{y|i+\frac12,j,k}-  \mu_0 J^{n+\frac12}_{x|i+\frac12,j,k} \right)
 \\
 ={}& 2E^{n}_{x|i+\frac12,j,k} - E^{n}_{x|i+\frac12,j,k} - c^2\Delta t \mu_0 J^{n+\frac12}_{x|i+\frac12,j,k} \\
 \nonumber
  &+ c^2\Delta t \mu_0 J^{n-\frac12}_{x|i+\frac12,j,k} - c^2\Delta t \mu_0 J^{n-\frac12}_{x|i+\frac12,j,k}
 \\
 \nonumber
 & + c^2\Delta t  D_y \left( B^{n-\frac12}_{z|i+\frac12,j,k} + \Delta t \left( D^*_y E^n_{x|i+\frac12,j,k} - D^*_x E^n_{y|i+\frac12,j,k} \right) \right)  \\
 \nonumber
 & - c^2\Delta t  D_z \left( B^{n-\frac12}_{y|i+\frac12,j,k} + \Delta t \left( D^*_x E^n_{z|i+\frac12,j,k} - D^*_z E^n_{x|i+\frac12,j,k} \right) \right)
 \\
 ={} & 2E^{n}_{x|i+\frac12,j,k} -E^{n-1}_{x|i+\frac12,j,k} -c^2\Delta t^2 \mu_0 D_t J^{n}_{x|i+\frac12,j,k}
 \\ \nonumber
 & + c^2\Delta t^2  D_y \left( D^*_y E^n_{x|i+\frac12,j,k} - D^*_x E^n_{y|i+\frac12,j,k} \right)  \\
 \nonumber
 & - c^2\Delta t^2  D_z \left( D^*_x E^n_{z|i+\frac12,j,k} - D^*_z E^n_{x|i+\frac12,j,k} \right)\,.
\end{align}
Which we can write as
\begin{align}
\frac{1}{c^2} D_t^2 E^{n}_{x|i+\frac12,j,k} &= \left( \vv{D}\cdot\vv{D}^* \right)E^{n}_{x|i+\frac12,j,k} - D_x^* \left( \vv{D}\cdot \vv{E}^n_{i+\frac12,j,k} \right) -\mu_0 D_t J^{n}_{x|i+\frac12,j,k}
\\
&= \left( \vv{D}\cdot\vv{D}^* \right)E^{n}_{x|i+\frac12,j,k} - \frac{1}{\varepsilon_0} D_x^* \left( \rho_E  \right)^n_{i+\frac12,j,k} -\mu_0 D_t J^{n}_{x|i+\frac12,j,k}
\end{align}
or
\begin{align}
 \label{eq:numerical_wave_equation_electric_charged}
 \frac{1}{c^2} D_t^2 \vv{E}^{n}_{i,j,k} &= \Delta^*\vv{E}^{n}_{i,j,k} -  \frac{1}{\varepsilon_0} \vv{D}^* \left( \rho_E  \right)^n_{i,j,k} -\mu_0 D_t \vv{J}^{n}_{i,j,k}\,.
\end{align}
This means that the electric field behaves according to the current density $\vv{J}$ and the charge density $\rho_E$ and we really have to make sure that $\rho_E$ is equal to the external charge density.

\section{Symmetries of the extended stencil}
\label{sec:betasymmetry}
Using
\begin{align}
 s_x^2 = \frac{1}{\Delta x^2} \sin^2\left( \frac12 k_x\Delta x \right) = \frac{1}{2\Delta x^2}\left( 1-\cos(k_x\Delta x) \right) = \frac{1}{2\Delta x^2}\left( 1-c_x \right)
\end{align}
and analogue expressions for $s_y^2$ and $s_z^2$ we can write the right hand side of \Eqref{eq:dispersion_relation} as
\begin{align}
 s_\omega^2 ={}& \sum_{i=x,y,z}  s_i^2A_i \\
 ={}& \sum_{i=x,y,z}  s_i^2 (\alpha_i + \delta_i (1+2c_i)) + \sum_{\substack{i,j=x,y,z\\i\neq j}} \frac{1}{\Delta x_i^2}\left( 1-c_i \right) \beta_{ij} c_j
 \\ ={}& \sum_{i=x,y,z}  s_i^2 \left(1-2\sum_{\substack{j=x,y,z\\j\neq i}}\beta_{ij}-3\delta_i + \delta_i (1+2c_i)\right)
 \\\nonumber
 &+ \sum_{\substack{i,j=x,y,z\\i\neq j}}  \frac{\beta_{ij}}{\Delta x_i^2}  \left( c_j - c_i c_j \right)
 \\ ={}& \sum_{i=x,y,z}  s_i^2 \left(1 - 2\delta_i (1-c_i)\right)-\sum_{i=x,y,z}  \frac{1}{\Delta x_i^2}\left( 1-c_i \right)\left(\sum_{\substack{j=x,y,z\\j\neq i}}\beta_{ij}\right)
 \\\nonumber
 & + \sum_{\substack{i,j=x,y,z\\i\neq j}}  \frac{\beta_{ij}}{\Delta x_i^2}  \left( c_j - c_i c_j \right)
 \\ ={}& \sum_{i=x,y,z}  s_i^2 \left(1 - 4\Delta x_i^2 \delta_i s_i^2\right) + \sum_{\substack{i,j=x,y,z\\i\neq j}}  \frac{\beta_{ij}}{\Delta x_i^2}  \left( -1+c_i+c_j - c_i c_j \right)
 \\ ={}& \sum_{i=x,y,z}  s_i^2 \left(1 - 4\Delta x_i^2 \delta_i s_i^2\right) + 4\sum_{\substack{i,j=x,y,z\\i\neq j}} \Delta x_j^2 \beta_{ij} s_i^2 s_j^2\,.
\end{align}
The term
\begin{equation}
 \sum_{\substack{i,j=x,y,z\\i\neq j}} \Delta x_j^2 \beta_{ij} s_i^2 s_j^2  =  \sum_{\substack{i,j=x,y,z\\i\neq j}} \hat{\beta}_{ij} s_i^2s_j^2
\end{equation}
clearly only depends on the symmetric part of $\hat{\beta}_{ij}$, as the antisymmetric part cancels out when contracted with the symmetric matrix $s_i^2s_j^2$.

\section*{References}

\begin{thebibliography}{10}
\expandafter\ifx\csname url\endcsname\relax
  \def\url#1{\texttt{#1}}\fi
\expandafter\ifx\csname urlprefix\endcsname\relax\def\urlprefix{URL }\fi
\expandafter\ifx\csname href\endcsname\relax
  \def\href#1#2{#2} \def\path#1{#1}\fi

\bibitem{inan2011numerical}
U.~S. Inan, R.~A. Marshall, {Numerical electromagnetics: the FDTD method},
  Cambridge University Press, 2011.

\bibitem{Birdsall:1991}
C.~K. Birdsall, A.~B. Langdon, {Plasma physics via computer simulation}, IOP
  Publishing Ltd., Bristol, 1991.

\bibitem{Yee1966}
K.~S. Yee, {Numerical solution of initial boundary value problems involving
  Maxwell's equations in isotropic media}, IEEE Trans. Antennas Propag. 14~(3)
  (1966) 302--307.
\newblock \href {http://dx.doi.org/10.1109/TAP.1966.1138693}
  {\path{doi:10.1109/TAP.1966.1138693}}.

\bibitem{Valcarce2010}
A.~Valcarce, H.~Song, J.~Zhang, {Characterization of the numerical group
  velocity in yee's FDTD grid}, IEEE Trans. Antennas Propag. 58~(12) (2010)
  3974--3982.
\newblock \href {http://dx.doi.org/10.1109/TAP.2010.2078453}
  {\path{doi:10.1109/TAP.2010.2078453}}.

\bibitem{Godfrey1974}
B.~B. Godfrey, {Numerical Cherenkov instabilities in electromagnetic particle
  codes}, J. Comput. Phys. 15~(4) (1974) 504--521.
\newblock \href {http://dx.doi.org/10.1016/0021-9991(74)90076-X}
  {\path{doi:10.1016/0021-9991(74)90076-X}}.

\bibitem{Godfrey:2014vxa}
B.~B. Godfrey, {Review and recent advances in PIC modeling of relativistic
  beams and plasmas}, in: AIP Conf. Proc., 1777, 2016, p. 020004.
\newblock \href {http://dx.doi.org/10.1063/1.4965593}
  {\path{doi:10.1063/1.4965593}}.

\bibitem{Liu1997}
Q.~Liu, {The pseudospectral time-domain (PSTD) method: a new algorithm for
  solutions of Maxwell's equations}, in: IEEE Antennas Propag. Soc. Int. Symp.
  1997. Dig., 2, IEEE, 1997, pp. 122--125.
\newblock \href {http://dx.doi.org/10.1109/APS.1997.630102}
  {\path{doi:10.1109/APS.1997.630102}}.

\bibitem{Lifschitz2009}
A.~F. Lifschitz, X.~Davoine, E.~Lefebvre, J.~Faure, C.~Rechatin, V.~Malka,
  {Particle-in-Cell modelling of laser-plasma interaction using Fourier
  decomposition}, J. Comput. Phys. 228~(5) (2009) 1803--1814.
\newblock \href {http://dx.doi.org/10.1016/j.jcp.2008.11.017}
  {\path{doi:10.1016/j.jcp.2008.11.017}}.

\bibitem{Vay2013}
J.-L. Vay, I.~Haber, B.~B. Godfrey, {A domain decomposition method for
  pseudo-spectral electromagnetic simulations of plasmas}, J. Comput. Phys. 243
  (2013) 260--268.
\newblock \href {http://dx.doi.org/10.1016/j.jcp.2013.03.010}
  {\path{doi:10.1016/j.jcp.2013.03.010}}.

\bibitem{Yu2015}
P.~Yu, X.~Xu, A.~Tableman, V.~K. Decyk, F.~S. Tsung, F.~Fiuza, A.~Davidson,
  J.~Vieira, R.~A. Fonseca, W.~Lu, L.~O. Silva, W.~B. Mori, {Mitigation of
  numerical Cerenkov radiation and instability using a hybrid finite
  difference-FFT Maxwell solver and a local charge conserving current deposit},
  Comput. Phys. Commun. 197 (2015) 144--152.
\newblock \href {http://arxiv.org/abs/arXiv:1502.01376v1}
  {\path{arXiv:arXiv:1502.01376v1}}, \href
  {http://dx.doi.org/10.1016/j.cpc.2015.08.026}
  {\path{doi:10.1016/j.cpc.2015.08.026}}.

\bibitem{Andriyash2016}
I.~A. Andriyash, R.~Lehe, A.~F. Lifschitz, {Laser-plasma interactions with a
  Fourier-Bessel Particle-in-Cell method}, Phys. Plasmas 23~(2) (2016) 033110.
\newblock \href {http://arxiv.org/abs/1512.09289} {\path{arXiv:1512.09289}},
  \href {http://dx.doi.org/10.1063/1.4943281} {\path{doi:10.1063/1.4943281}}.

\bibitem{Lehe2016}
R.~Lehe, M.~Kirchen, I.~A. Andriyash, B.~B. Godfrey, J.-L. Vay, {A spectral,
  quasi-cylindrical and dispersion-free Particle-In-Cell algorithm}, Comput.
  Phys. Commun. 203 (2016) 66--82.
\newblock \href {http://arxiv.org/abs/1507.04790} {\path{arXiv:1507.04790}},
  \href {http://dx.doi.org/10.1016/j.cpc.2016.02.007}
  {\path{doi:10.1016/j.cpc.2016.02.007}}.

\bibitem{Greenwood2002}
A.~D. Greenwood, K.~L. Cartwright, E.~A. Baca, J.~W. Luginsland, {On the use of
  FDTD to simulate systems with charged particles}, in: IEEE Antennas Propag.
  Soc. Int. Symp. (IEEE Cat. No.02CH37313), Vol.~3, IEEE, 2002, pp. 268--271.
\newblock \href {http://dx.doi.org/10.1109/APS.2002.1018207}
  {\path{doi:10.1109/APS.2002.1018207}}.

\bibitem{Wang2003}
S.~Wang, F.~Teixeira, {Dispersion-relation-preserving FDTD algorithms for
  large-scale three-dimensional problems}, IEEE Trans. Antennas Propag. 51~(8)
  (2003) 1818--1828.
\newblock \href {http://dx.doi.org/10.1109/TAP.2003.815435}
  {\path{doi:10.1109/TAP.2003.815435}}.

\bibitem{Greenwood2004}
A.~D. Greenwood, K.~L. Cartwright, J.~W. Luginsland, E.~A. Baca, {On the
  elimination of numerical Cerenkov radiation in PIC simulations}, J. Comput.
  Phys. 201~(2) (2004) 665--684.
\newblock \href {http://dx.doi.org/10.1016/j.jcp.2004.06.021}
  {\path{doi:10.1016/j.jcp.2004.06.021}}.

\bibitem{Vay2011}
J.-L. Vay, C.~G.~R. Geddes, E.~Cormier-Michel, D.~Grote, {Numerical methods for
  instability mitigation in the modeling of laser wakefield accelerators in a
  Lorentz-boosted frame}, J. Comput. Phys. 230~(15) (2011) 5908--5929.
\newblock \href {http://dx.doi.org/10.1016/j.jcp.2011.04.003}
  {\path{doi:10.1016/j.jcp.2011.04.003}}.

\bibitem{Webb2016}
S.~D. Webb, D.~T. Abell, N.~M. Cook, D.~L. Bruhwiler, {A Spectral Symplectic
  Algorithm for Cylindrical Electromagnetic Plasma Simulations}, ArXiv\href
  {http://arxiv.org/abs/1609.05095} {\path{arXiv:1609.05095}}.

\bibitem{Yanenko2012}
N.~N. Yanenko, {The Method of Fractional Steps: The Solution of Problems of
  Mathematical Physics in Several Variables}, Springer Berlin Heidelberg,
  Berlin, Heidelberg, 1971.
\newblock \href {http://dx.doi.org/10.1007/978-3-642-65108-3}
  {\path{doi:10.1007/978-3-642-65108-3}}.

\bibitem{LichtersLPIC}
R.~E.~W. Pfund, R.~Lichters, J.~Meyer-ter Vehn, {LPIC++ a parallel
  one-dimensional relativistic electromagnetic Particle-In-Cell code for
  simulating laser-plasma-interaction}, in: AIP Conf. Proc., 426, 1998, pp.
  141--146.
\newblock \href {http://dx.doi.org/10.1063/1.55199}
  {\path{doi:10.1063/1.55199}}.

\bibitem{Lele1992}
S.~K. Lele, {Compact finite difference schemes with spectral-like resolution},
  J. Comput. Phys. 103~(1) (1992) 16--42.
\newblock \href {http://arxiv.org/abs/fld.1} {\path{arXiv:fld.1}}, \href
  {http://dx.doi.org/10.1016/0021-9991(92)90324-R}
  {\path{doi:10.1016/0021-9991(92)90324-R}}.

\bibitem{Georgakopoulos2002}
S.~V. Georgakopoulos, C.~R. Birtcher, C.~A. Balanis, R.~A. Renaut,
  {Higher-order finite-difference schemes for electromagnetic radiation,
  scattering, and penetration, Part 2: Applications}, IEEE Antennas Propag.
  Mag. 44~(2) (2002) 92--101.
\newblock \href {http://dx.doi.org/10.1109/MAP.2002.1003639}
  {\path{doi:10.1109/MAP.2002.1003639}}.

\bibitem{Panaretos2006}
A.~Panaretos, J.~Aberle, R.~D{\'{i}}az, {A Three-Dimensional Finite-Difference
  Time-Domain Scheme Based on a Transversely Extended-Curl Operator}, IEEE
  Trans. Microw. Theory Tech. 54~(12) (2006) 4237--4246.
\newblock \href {http://dx.doi.org/10.1109/TMTT.2006.885900}
  {\path{doi:10.1109/TMTT.2006.885900}}.

\bibitem{Finkelstein2007}
B.~Finkelstein, R.~Kastner, {Finite difference time domain dispersion reduction
  schemes}, J. Comput. Phys. 221~(1) (2007) 422--438.
\newblock \href {http://dx.doi.org/10.1016/j.jcp.2006.06.016}
  {\path{doi:10.1016/j.jcp.2006.06.016}}.

\bibitem{Karkkainen2006}
M.~K{\"{a}}rkk{\"{a}}inen, E.~Gjonaj, T.~Lau, T.~Weiland, {Low-Dispersion Wake
  Field Calculation Tools}, in: Proc. ICAP 2006, Vol.~1, Chamonix, France,
  2006, p.~35.

\bibitem{Pukhov1999}
A.~Pukhov, {Three-dimensional electromagnetic relativistic particle-in-cell
  code VLPL (Virtual Laser Plasma Lab)}, J. Plasma Phys. 61~(3) (1999)
  425--433.

\bibitem{Cowan2013}
B.~M. Cowan, D.~L. Bruhwiler, J.~R. Cary, E.~Cormier-Michel, C.~G.~R. Geddes,
  {Generalized algorithm for control of numerical dispersion in explicit
  time-domain electromagnetic simulations}, Phys. Rev. Accel. Beams 16~(4)
  (2013) 041303.
\newblock \href {http://dx.doi.org/10.1103/PhysRevSTAB.16.041303}
  {\path{doi:10.1103/PhysRevSTAB.16.041303}}.

\bibitem{Lehe2013}
R.~Lehe, A.~F. Lifschitz, C.~Thaury, V.~Malka, X.~Davoine, {Numerical growth of
  emittance in simulations of laser-wakefield acceleration}, Phys. Rev. Accel.
  Beams 16~(2) (2013) 021301.
\newblock \href {http://dx.doi.org/10.1103/PhysRevSTAB.16.021301}
  {\path{doi:10.1103/PhysRevSTAB.16.021301}}.

\bibitem{Nuter2014}
R.~Nuter, M.~Grech, P.~{Gonzalez De Alaiza Martinez}, G.~Bonnaud,
  E.~D'Humi{\`{e}}res, {Maxwell solvers for the simulations of the laser-matter
  interaction}, Eur. Phys. J. D 68~(6) (2014) 1--9.
\newblock \href {http://dx.doi.org/10.1140/epjd/e2014-50162-y}
  {\path{doi:10.1140/epjd/e2014-50162-y}}.

\bibitem{Kraft1988}
D.~Kraft, {A software package for sequential quadratic programming},
  Forschungsbericht. Dtsch. Forschungs- und Versuchsanstalt f{\"{u}}r Luft- und
  Raumfahrt, DFVLR.

\bibitem{SciPy}
E.~Jones, T.~Oliphant, P.~Petersen, et~al.,
  \href{http://www.scipy.org/}{{SciPy: Open Source Scientific Tools for
  Python}} (2001--).
\newline\urlprefix\url{http://www.scipy.org/}

\bibitem{optimization_code}
A.~Blinne, D.~Schinkel,
  \href{https://github.com/Ablinne/optimize-stencil}{Optimize stencil} (2017).
\newline\urlprefix\url{https://github.com/Ablinne/optimize-stencil}

\bibitem{ArberEPOCH}
T.~D. Arber, K.~Bennett, C.~S. Brady, A.~Lawrence-Douglas, M.~G. Ramsay, N.~J.
  Sircombe, P.~Gillies, R.~G. Evans, H.~Schmitz, a.~R. Bell, C.~P. Ridgers,
  {Contemporary particle-in-cell approach to laser-plasma modelling}, Plasma
  Phys. Control. Fusion 57~(11) (2015) 113001.
\newblock \href {http://dx.doi.org/10.1088/0741-3335/57/11/113001}
  {\path{doi:10.1088/0741-3335/57/11/113001}}.

\bibitem{Hunter2007}
J.~D. Hunter, {Matplotlib: A 2D Graphics Environment}, Comput. Sci. Eng. 9~(3)
  (2007) 90--95.
\newblock \href {http://arxiv.org/abs/0402594v3} {\path{arXiv:0402594v3}},
  \href {http://dx.doi.org/10.1109/MCSE.2007.55}
  {\path{doi:10.1109/MCSE.2007.55}}.

\bibitem{Perez2007}
F.~Perez, B.~E. Granger, {IPython: A System for Interactive Scientific
  Computing}, Comput. Sci. Eng. 9~(3) (2007) 21--29.
\newblock \href {http://dx.doi.org/10.1109/MCSE.2007.53}
  {\path{doi:10.1109/MCSE.2007.53}}.

\bibitem{Esirkepov2001}
T.~Z. Esirkepov, {Exact charge conservation scheme for Particle-in-Cell
  simulation with an arbitrary form-factor}, Comput. Phys. Commun. 135~(2)
  (2001) 144--153.
\newblock \href {http://arxiv.org/abs/9901047} {\path{arXiv:9901047}}, \href
  {http://dx.doi.org/10.1016/S0010-4655(00)00228-9}
  {\path{doi:10.1016/S0010-4655(00)00228-9}}.

\end{thebibliography}

\end{document}